\documentclass[twocolumn,showpacs,amsmath,amssymb,aps,pra,floatfix]{revtex4-1}

\usepackage{float}
\usepackage{lipsum}
\usepackage{epsfig}
\usepackage{amsmath}
\usepackage{amssymb}
\usepackage{bm}
\usepackage{graphicx}
\usepackage{color}
\usepackage{graphicx,rotating}

\begin{document}
\title{Fields of an ultrashort tightly focused radially polarized laser pulse in a linear response plasma}

\author{Yousef I. Salamin}
\affiliation{Department of Physics, American University of Sharjah, POB 26666, Sharjah, United Arab Emirates}

\date{\today}

\begin{abstract}

Analytic expressions for the fields of a radially polarized, ultrashort and tightly focused laser pulse propagating in a linear-response plasma are derived and discussed. The fields are obtained from solving the inhomogenous wave equations for the vector and scalar potentials, linked by the Lorenz gauge, in a plasma background. First the scalar potential is eliminated using the gauge condition, then the  vector potential is synthesized from Fourier components of an initial uniform distribution of wavenumbers, and the inverse Fourier transformation is carried out term-by-term in a truncated series (finite sum). The zeroth-order term in, for example, the axial electric field component is shown to model a pulse much better than its widely used paraxial approximation counterpart. Some of the propagation characteristics of the fields are discussed and all fields are shown to have manifestly the expected limits for propagation in a vacuum. 

\end{abstract}

\maketitle

\section{Introduction}

Present-day high power laser systems, which find applications in many fields, are mostly of the pulsed type \cite{vulkan,eli}. For example, attosecond science \cite{brabec1}, femtosecond x-ray generation \cite{corde}, and laser acceleration of particles for medical and industrial applications, all require ultrashort and tightly focused laser pulses \cite{prat,esarey0,benedetti,wang}. The term {\it ultrashort} in this work refers to the axial extension in space, $L$, being small compared to the Rayleigh length $z_r$ (or even a wavelength $\lambda$). Relation to the temporal pulse length may be taken loosely to be $L \sim c\tau_0$. In the term {\it tightly focused}, on the other hand, the reference is to the transverse extension being small compared to the waist radius at focus $w_0$. Theoretical investigations utilizing these pulses rely mostly on numerical simulations of the complex interactions of the radiation field with material particles. While the numerical investigations do well in handling complex situations, the equations and further analytic work can paint a much clearer intuitive picture.

This paper derives expressions for, and investigates, mostly analytically, the electric and magnetic fields, and the propagation characteristics in a plasma of linear response, of an ultrashort and tightly focused laser pulse of the radially polarized variety. A radially polarized laser beam can be focused to a tighter spot than would a linearly polarized one \cite{dorn}. This is important for laser acceleration and other applications for which the formation of filaments (self-focusing) and the reshaping of the laser profile are needed. The power and intensity levels of radially polarized light are still substantially lower than what is currently available from the linearly polarized sources, but that is bound to change as radially polarized light continues to find new applications \cite{navarro,yu}. For example, the radially polarized electric field component helps to trap charged particles, while the strong axial component works to accelerate them, in a laser-accelerator setup \cite{salamin2,salamin3,salamin4,varin2,lin}. For many applications, including laser-acceleration, source of the electrons or ions may be a rarified gas or a slowly expanding cluster of atoms. Irradiation of the source with the laser pulse, produces a low-density plasma. Components of the resulting plasma then serve as sources of charge and current densities which should be taken into account when Maxwell equations, or the wave equations based upon them, are solved for the fields.

Our analytic investigations will go along the following lines. First, an appropriate model is adopted for the sources in the wave equations satisfied by the scalar and vector potentials. Second, the Lorenz gauge is used to express the scalar potential in terms of the vector potential, thus rendering the need to solve the wave equation of the scalar potential unnecessary. Third, the wave equation of the vector potential is solved using a new set of coordinates, and employing a Fourier transform in one of those coordinates. Fourth, the inverse Fourier transformation is then carried out by introducing a power series expansion and carrying out the integrations term by term. Fifth, the zeroth-order vector potential, which corresponds to the first term in the above-mentioned power series, is used to derive the zeroth-order electric and magnetic fields of the pulse.

This work follows the steps outlined in Refs. \cite{salaminPRA92-1} and \cite{salaminPRA92-2}, with the important added feature of propagation in a plasma, rather than a vacuum. There is also an important difference, namely, here the initial distribution is centered about $k=0$. This avoids having to worry about the appearance of an extra phase term upon inverse Fourier transformation. The analysis presented here also benefits from the approach introduced some twenty years ago by Esarey {\it et al.} \cite{esarey}. It should be borne in mind, though, that the work of Esarey {\it et al.}  presents fields for a linearly polarized pulse, uses only a vector potential, and employs Fourier components having an initial Gaussian distribution of wavenumbers. By contrast, this work is for a radially polarized pulse, employs a scalar as well as a vector potential, and uses a uniform distribution of wavenumbers.

In Sec. \ref{sec:equations} the main working equations will be briefly reviewed, which will serve as a basis for the subsequent investigations. The wave equation satisfied by the vector potential in the plasma will be solved in Sec. \ref{sec:solution}, using a Fourier transform approach. A truncated series will be arrived at for the vector potential in this section, and a program will be described for analytic and numerical evaluation of the series to any desired order of truncation. This is followed, in Sec. \ref{sec:zeroth}, by a rigorous derivation of the zeroth-order electric and magnetic fields, from the zeroth-order vector potential. A dispersion relation is derived, and some of the pulse propagation characteristics are described in in Sec. \ref{sec:prop}. Finally, some concluding remarks will be given in Sec. \ref{sec:conc}.

\section{The main working equations}\label{sec:equations}

In the presence of sources, Maxwell's equations for the electric and magnetic fields $\bm{E}$ and $\bm{B}$, respectively, are (SI units)
\begin{eqnarray}
 \label{maxwell1} \bm{\nabla}\times\bm{H} &=& \bm{J}+\frac{\partial\bm{D}}{\partial t},\\
 \label{maxwell2} \bm{\nabla}\times\bm{E} &=& -\frac{\partial\bm{B}}{\partial t},\\
 \label{maxwell3} \bm{\nabla}\cdot\bm{B} &=& 0,\\
 \label{maxwell4} \bm{\nabla}\cdot\bm{D} &=& \varrho,
\end{eqnarray}
where, assuming a medium of linear response, $\bm{D}=\varepsilon_0\bm{E}$ and $\bm{B}=\mu_0\bm{H}$, and $\varepsilon_0$ and $\mu_0$ being the permittivity and permeability, respectively, of the vacuum. The sources in Eqs. (\ref{maxwell1}) and (\ref{maxwell4}) are the current and charge densities, $\bm{J}$ and $\varrho$, respectively, which are related through the continuity equation
\begin{equation}\label{cont}
  \bm{\nabla}\cdot\bm{J}+\frac{\partial\varrho}{\partial t} = 0.
\end{equation}
In terms of the scalar and vector potentials $\Phi$ and $\bm{A}$, respectively, Maxwell's equations are equivalent to the wave equations \cite{jackson}
\begin{eqnarray}
 \label{weA} \left(\nabla^2-\frac{1}{c^2}\frac{\partial^2}{\partial t^2}\right)\bm{A} &=& -\mu_0\bm{J},\\
 \label{wePhi} \left(\nabla^2-\frac{1}{c^2}\frac{\partial^2}{\partial t^2}\right)\Phi &=& -\frac{\varrho}{\varepsilon_0},
\end{eqnarray}
where $c$ is the speed of light in vacuum, provided the Lorenz gauge
\begin{equation}\label{lorenz}
    \bm{\nabla}\cdot\bm{A}+\frac{1}{c^2}\frac{\partial\Phi}{\partial t} = 0,
\end{equation}
is simultaneously satisfied \cite{jackson}.

Equations (\ref{maxwell1})--(\ref{lorenz}) are general. In the present work, however, we apply them to the propagation of a laser pulse in a plasma, in which the plasma response to the laser fields must be accounted for through $\varrho$ and $\bm{J}$. For example, in Eq. (\ref{weA}) the term $-\mu_0\bm{J}$ gets replaced \cite{esarey} effectively by $k_p^2\bm{A}$, where $k_p = \omega_p/c$ is an effective plasma wavenumber, and $\omega_p=\sqrt{n_0e^2/(m\varepsilon_0)}$ is the frequency of plasma oscillations, in which $n_0$ is the number density of the ambient electrons and $m$ and $-e$ are the mass and charge, respectively, of the electron. In general, the functional dependence of $k_p$ on $\bm{A}$ is highly nonlinear \cite{esarey2,esarey3,esarey4}. However, attention in this work will be limited to the case of a linear plasma response, for which $k_p$ is a constant. This is often considered a valid assumption in the non-relativistic interaction regime, for which $|e\bm{A}/(mc^2)|^2 \ll 1$. On the other hand, insertion of $\bm{J}$ from $-\mu_0\bm{J} \approx k_p^2\bm{A}$ into the continuity equation, using the Lorenz condition, and integrating the result formally, gives $-\varrho/\varepsilon_0 \approx k_p^2\Phi$. Under these conditions, Eqs. (\ref{weA}) and (\ref{wePhi}) become
\begin{eqnarray}
 \label{weA1} \left(\nabla^2-\frac{1}{c^2}\frac{\partial^2}{\partial t^2}-k_p^2\right)\bm{A} &=& 0,\\
 \label{wePhi1} \left(\nabla^2-\frac{1}{c^2}\frac{\partial^2}{\partial t^2}-k_p^2\right)\Phi &=& 0.
\end{eqnarray}

For a linearly polarized vector potential, Eqs. (\ref{weA1}) and (\ref{wePhi1}) have the same structure {\it mathematically} and, thus, it will be postulated that their solutions can have the same formal {\it mathematical} structure. Specifically, the following forms will be assumed \cite{mcdonald}
\begin{eqnarray}
 \label{ansatzA} \bm{A} &=& \hat{z} a_0 a(x,y,z,t) e^{ik_0(z-ct)},\\
 \label{ansatzPhi} \Phi &=& \phi_0\phi(x,y,z,t) e^{ik_0(z-ct)},
\end{eqnarray}
where $\hat{z}$ is a unit vector in the propagation direction of the laser pulse, $a_0$ and $\phi_0$ are constant complex amplitudes, and $k_0$ is a central wave number, the latter corresponding to a central laser angular frequency $\omega_0 = ck_0$.

The electromagnetic fields, and propagation characteristics in the plasma, of an ultrashort, tightly focused laser pulse of the radially polarized variety, may be obtained from the vector and scalar potentials (\ref{ansatzA}) and (\ref{ansatzPhi}). It can further be shown \cite{li,salaminPRA92-1,salaminPRA92-2} that the Lorenz gauge condition (\ref{lorenz}) leads to
\begin{equation}\label{R}
  \Phi = \frac{c^2(\bm{\nabla}\cdot\bm{A})}{R};\quad R = ick_0-\frac{1}{a}\frac{\partial a}{\partial t}.
\end{equation}
Thus, one needs only to worry about finding a solution to Eq. (\ref{weA1}). Once an analytic expression for $\bm{A}$ has been found, the electric and magnetic fields of the pulse will readily be obtained from
\begin{equation}\label{EB}
  \bm{E} = -\bm{\nabla}\Phi-\frac{\partial\bm{A}}{\partial t}; \quad \text{and}\quad \bm{B} = \bm{\nabla}\times \bm{A}.
\end{equation}

In cylindrical coordinates, a vector potential polarized axially, i.e., along the direction of propagation of the pulse, the $z-$axis, will eventually result in a two-component $\bm{E}$ field, one component polarized radially and the other axially, whereas the resulting $\bm{B}$ field will have a single component polarized azimuthally \cite{mcdonald}.

\section{Solution to the wave equation}\label{sec:solution}

The wave equation (\ref{weA1}) can most conveniently be solved in terms of the set of coordinates
\begin{equation}\label{variables}
  \rho = \frac{\sqrt{x^2+y^2}}{w_0};\quad \eta = \frac{z+ct}{2};\quad \zeta = z-ct.
\end{equation}
In terms of the new variables, Eq. (\ref{weA1}) transforms into
\begin{equation}\label{Aeq}
  \left(\frac{1}{\rho}\frac{\partial}{\partial\rho}\rho\frac{\partial}{\partial\rho}
    +2w_0^2\frac{\partial^2}{\partial\eta\partial\zeta}-k_p^2w_0^2\right)\bm{A} = 0,
\end{equation}
and the vector potential (\ref{ansatzA}) becomes
\begin{equation}\label{ansatz}
  \bm{A}(\rho,\eta,\zeta) = \hat{z}a_0a(\rho,\eta,\zeta)e^{ik_0\zeta}.
\end{equation}
Note that substitution of (\ref{ansatz}) into (\ref{Aeq}) gives an equation for $a(\rho, \eta, \zeta)$. Then, employing the Fourier transform pair
\begin{eqnarray}
 \label{fourier} a(\rho, \eta, \zeta) &=& \frac{1}{\sqrt{2\pi}}\int_{-\infty}^{\infty}a_k(\rho, \eta,k)e^{ik\zeta} dk,\\
\label{invfourier} a_k(\rho, \eta, k) &=& \frac{1}{\sqrt{2\pi}}\int_{-\infty}^{\infty}a(\rho, \eta,\zeta)e^{-ik\zeta} d\zeta,
\end{eqnarray}
will turn that equation into
\begin{equation}\label{akeq}
  \left(\frac{1}{\rho}\frac{\partial}{\partial\rho}\rho\frac{\partial}{\partial\rho}
    +4iz_{rk}\frac{\partial}{\partial\eta}-k_p^2w_0^2\right)a_k=0,
\end{equation}
for each Fourier component $a_k$. In (\ref{akeq}) the quantity $z_{rk} = (k+k_0)w_0^2/2$, which reduces to the Rayleigh length $z_r=k_0w_0^2/2$, for $k=0$.

Equation (\ref{akeq}) admits the following exact solution, which may readily be verified by direct substitution
\begin{equation}\label{exact}
  a_k = \frac{f_k}{1+i\alpha_k} \exp\left[-\frac{\rho^2}{1+i\alpha_k}-\frac{i\alpha_kk_p^2w_0^2}{4}\right],
\end{equation}
in which $\alpha_k = \eta/z_{rk}$, and $f_k$ is an appropriate function of $k$, but is independent of $\zeta$. Note that if the focus of the pulse is taken initially (at $t=0$) to be at the origin of the coordinate system (with $z=0$) its initial position will be at $\eta=0$ in the new system of coordinates. Hence, Eq. (\ref{exact}) yields $a_k(\rho,0,k) = f_k\exp(-\rho^2)$. Inverse Fourier transformation then gives $a(\rho,0,\zeta) = f(\zeta)\exp(-\rho^2)$, a Gaussian in the transverse coordinate. Furthermore, $f(\zeta)$ will be the Fourier transform of $f_k$, serving in this way as an initial axial profile for the pulse emitted at $t=0$.

A plausible choice for $f_k$ will be key to determining the appropriate initial axial profile $f(\zeta)$. The simplest would be a uniform distribution of wavenumbers, given by the square function  \cite{signal,shift}
\begin{equation}\label{fk0}
  f_k = \left\{
           \begin{array}{ll}
             \frac{\sqrt{2\pi}}{\Delta k}, & \hbox{$|k|\leq\frac{\Delta k}{2}$;} \\
             $0$, & \hbox{\text{elsewhere.}}
           \end{array}
         \right.
\end{equation}
This function is displayed graphically in Fig. 1(a) and the square modulus of its Fourier transform
\begin{equation}\label{envelope}
  f(\zeta) = \frac{1}{\sqrt{2\pi}}\int_{-\infty}^{\infty} f_k e^{ik\zeta}dk =  \frac{\sin(\zeta\Delta k/2)}{\zeta\Delta k/2},
\end{equation}
the sinc function, sinc$(\zeta\Delta k/2)$, is shown in Fig. 1(b). From Fig. 1(b) it seems appropriate to take, for the {\it initial} axial length of the pulse, the quantity $L=\Delta\zeta=2\pi/\Delta k$, which is roughly the same as the full-width-at-half-maximum (FWHM). This is also plausible because $z=\zeta$, at $t=0$. Equation (\ref{fk0}) for the initial wavenumber distribution (uniform, centered about $k_0$, and corresponding to a bandwidth $\Delta\omega=c\Delta k$) has recently been used by us in calculations of similar nature, albeit of the fields of radially and linearly polarized pulses propagating in a vacuum \cite{salaminPRA92-1,salaminPRA92-2}. Other wavenumber distribution functions have also been used elsewhere \cite{esarey,li}.

\begin{figure}
\includegraphics[width=4cm]{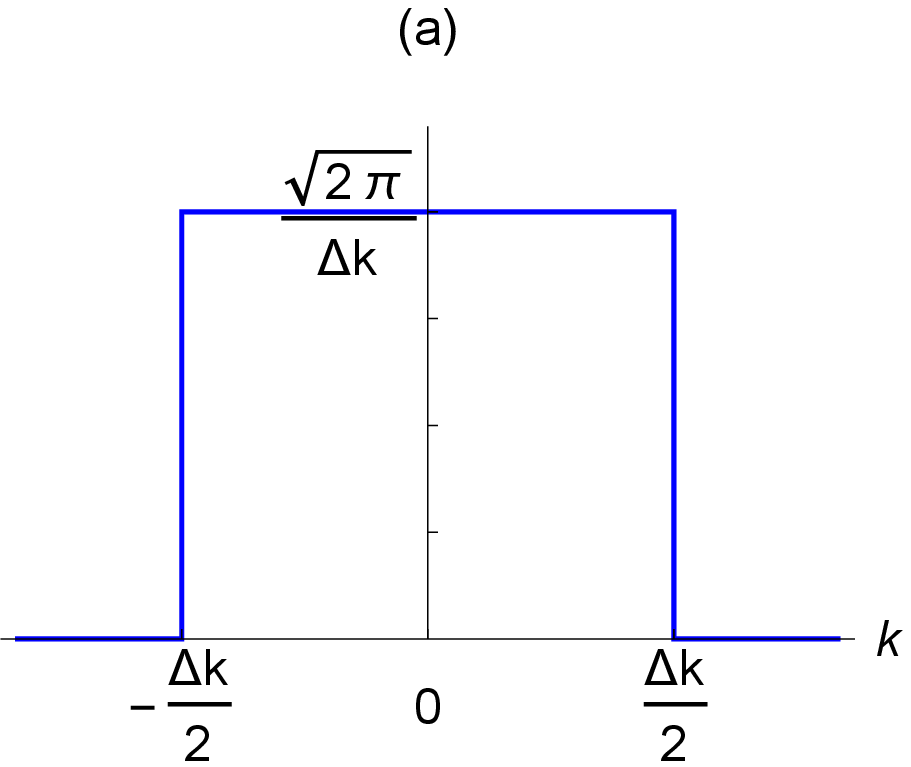}
\includegraphics[width=4cm]{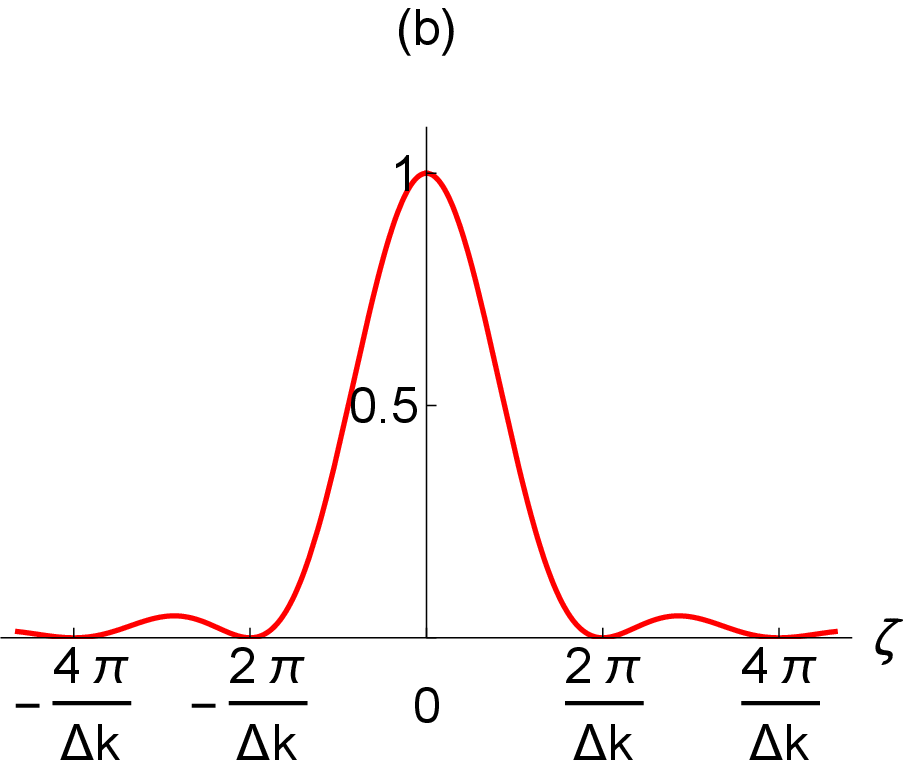}
\caption{(Color online) (a) The initial pulse wavenumber distribution function $f_k$. (a) The initial intensity envelope function in coordinate space $|f(\zeta)|^2$. Note that $f_k$ and $f(\zeta)$ constitute a Fourier transform pair. }
\label{fig1}
\end{figure}

Using (\ref{fk0}) the vector potential amplitude $a(\rho, \eta, \zeta)$ may now be synthesized from the Fourier components $a_k(\rho, \eta,k)$ according to Eq. (\ref{fourier}). In effect, one needs to evaluate
\begin{equation}\label{a}
   a(\rho,\eta,\zeta) = \frac{1}{\Delta k}\int_{-\frac{\Delta k}{2}}^{\frac{\Delta k}{2}} \psi_k e^{ik\zeta} dk,
\end{equation}
where
\begin{equation}\label{psik}
   \psi_k = \frac{1}{1+i\alpha_k} \exp\left[-\frac{\rho^2}{1+i\alpha_k}-\frac{i\alpha_k k_p^2w_0^2}{4}\right].
\end{equation}
No further attempts will be made to evaluate the remaining integral in (\ref{a}) in closed analytic form. Instead, $\psi_k$ will be viewed as a function of $k' \equiv k+k_0$, and will subsequently be power-series expanded about the central wavenumber $k_0$. Formally,
\begin{eqnarray}\label{taylor}
  \psi_k &=& \sum_{m=0}^\infty \frac{\partial^m\psi_k}{\partial k'^m}\bigg|_{k'=k_0} \frac{(k'-k_0)^m}{m!},\nonumber\\
         &=& \sum_{m=0}^\infty \psi_0^{(m)} \frac{k^m}{m!};\quad \psi_0^{(m)}(\rho,\eta)\equiv \frac{\partial^{m}\psi_k}{\partial k^m}\bigg|_{k=0}.
\end{eqnarray}
~~~\\
Now, using Eqs. (\ref{a})--(\ref{taylor}) in (\ref{ansatz}) gives
\begin{equation}\label{am0}
  A = \frac{a_0e^{ik_0\zeta}}{\Delta k}\sum_{m=0}^{\infty}\frac{\psi_0^{(m)}}{m!} \int_{-\frac{\Delta k}{2}}^{\frac{\Delta k}{2}} k^m e^{ik\zeta} dk.
\end{equation}
Note that, roughly speaking, the ratio of the term in the series (\ref{am1}) of order $m+1$ to that of order $m$ goes like $1/(m+1)\to0$, as $m\to\infty$. This is just a hint at convergence. A longer, but more rigorous, discussion of this issue may be found in \cite{salaminPRA92-1}. 

Admittedly, convergence of the series remains in doubt. The issue of convergence will be revisited shortly within a numerical analysis involving a few specific examples. On the other hand, note that the first few terms in the series are important in applications \cite{salPLA}. We, therefore, opt for truncating the series (\ref{am0}) at some finite order $n$. Thus, to order $n$, the vector potential takes on the following form
\begin{equation}\label{am1}
  	A^{(n)} = a_0e^{ik_0\zeta}\sum_{m=0}^{n} \frac{(-i)^m}{m!} \psi_0^{(m)} \frac{\partial^mf}{\partial\zeta^m},
\end{equation}
in which 
\begin{equation}
  \label{f} f(\zeta) = \frac{1}{\Delta k} \int_{-\frac{\Delta k}{2}}^{\frac{\Delta k}{2}}e^{ik\zeta} dk = \text{sinc}\left(\frac{\zeta\Delta k}{2}\right).
\end{equation}
Note also that $\text{sinc}(x) = j_0(x)$, the order-zero spherical Bessel function of the first kind. 

Working with Eq. (\ref{am1}) in which the spherical Bessel function of the first kind plays a central role is not the only option available. As it turns out, the integration in Eq. (\ref{am0}) can actually be carried out in closed analytic form. This gives the following alternative expression for the vector potential, in terms of incomplete gamma functions
\begin{eqnarray}\label{analytic}
	  A^{(n)} &=& \frac{a_0e^{ik_0\zeta}}{2\pi/L}\sum_{m=0}^{n} \frac{i^{m+1}}{m!\zeta^{m+1}} \psi_0^{(m)}\nonumber\\
	  & & \times\left[\Gamma\left(m+1,\frac{i\pi\zeta}{L}\right)-\Gamma\left(m+1,-\frac{i\pi\zeta}{L}\right)\right].
\end{eqnarray}

Unfortunately, the remaining sums in Eqs. (\ref{am1}) and (\ref{analytic}) are difficult to carry out analytically, and resort in working with those equations numerically seems inevitable. The good news is that one has to worry about evaluating only the zeroth-order term, or the first few terms at worst, as will be demonstrated shortly, and as has recently been alluded to elsewhere \cite{salaminPRA92-1,salaminPRA92-2}. 

For further analytic considerations, the first few expressions for $\psi_0^{(m)}$ are
\begin{equation}
 \label{p0} \psi_0^{(0)} = \frac{1}{p} \exp\left[-\frac{i\alpha k_p^2w_0^2}{4}-\frac{\rho^2}{p}\right],
 \end{equation}
 \begin{equation}
\psi_0^{(1)} = \frac{i\alpha}{k_0} \left[\frac{k_p^2w_0^2}{4}-\frac{\rho^2}{p^2}+\frac{1}{p}\right]\psi_0^{(0)},
\end{equation}
\begin{widetext}
\begin{equation}
 \label{p2} \psi_0^{(2)} = \frac{i\alpha}{k_0^2} \left[\frac{i\alpha k_p^4 w_0^4}{16}-\frac{\rho^4}{p^4}+\frac{\rho^2(4+\rho^2)}{p^3}+\frac{k_p^2w_0^2 \rho^2-4(1+\rho^2)}{2p^2}-\frac{k_p^2 w_0^2(1+\rho^2)}{2p} \right]\psi_0^{(0)},
\end{equation}
\begin{eqnarray}
 \label{p3} \psi_0^{(3)} &=& \frac{i\alpha}{k_0^3} \left\{-\frac{\alpha^2 k_p^6 w_0^6}{64}-\frac{\rho^6}{p^6}+\frac{2 \rho^6+9
   \rho ^4}{p^5}+\frac{3k_p^2 w_0^2\rho^4-4\rho^2(\rho^4+12\rho^2+18)}{4p^4}-\frac{3k_p^2w_0^2(2+\rho^2)\rho^2-6(\rho^4+4\rho ^2+2)}{2p^3}\right.\nonumber\\
    & &\left.+\frac{3 k_p^2 w_0^2\left[4\left(\rho^4+4\rho^2+2\right)-k_p^2w_0^2\rho^2\right]}{16p^2}+\frac{3k_p^4w_0^4(1+2\rho^2)}{16 p}-\frac{3k_p^4w_0^4(p+\rho^2)}{16}\right\}\psi_0^{(0)}.
\end{eqnarray}
\end{widetext}
In Eqs. (\ref{p0})--(\ref{p3})
\begin{equation}\label{P}
  p = 1+i\alpha,\quad \alpha = \frac{\eta}{z_r},\quad z_r = \frac{1}{2}k_0w_0^2.
\end{equation}
Note that the equations corresponding to (\ref{p0})--(\ref{p3}) in the case of propagation in vacuum \cite{salaminPRA92-1,salaminPRA92-2}, follow manifestly in the limit of $k_p\to0$. 

\begin{figure}
\includegraphics[width=8cm]{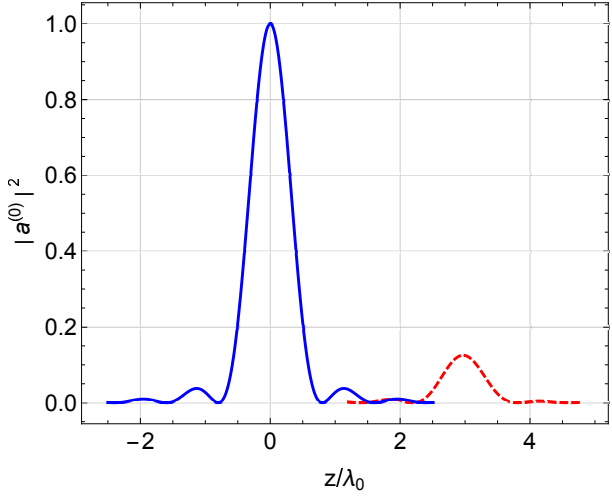}
\caption{(Color online) Snapshots of the square of the scaled zeroth-order vector potential amplitude $a^{(0)} = A^{(0)}/a_0$ on axis ($\rho = 0$). The blue line: at $t = 0$, and the red-dashed line: at $t = 10$ fs. For both cases, $w_0 = 0.6 \lambda_0$, $L = 0.8 \lambda_0$, $\varphi_0 = 0$, and the ambient electron density $n_0 = 5\times10^{25}$ m$^{-3}$.}
\label{fig2}
\end{figure}

In anticipation of the discussion, to be conducted in the next section, of the zeroth-order field components, square modulus of the zeroth-order vector potential is shown first in Fig. \ref{fig2} at times $t = 0$ and 10 fs. Note that the maximum of the square of the normalized vector potential $a^{(0)} \equiv A^{(0)}/a_0$ drops, after propagation for 10 fs, to about 10\% of its initial value at $t = 0$, due to diffraction. Furthermore, the centroid of the profile as a whole advances an axial distance $\Delta z\sim 3\lambda_0$, slightly less than  $ct$. More on this will be found in the next section.

\begin{figure}[b]
\includegraphics[width=8cm]{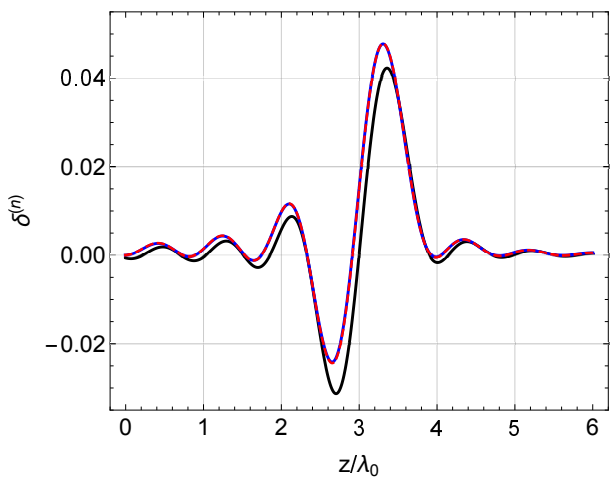}
\caption{(Color online) The uncertainties $\delta^{(n)}$, defined by Eq. (\ref{uncert}), sustained when only the zeroth-order term $|A^{(0)}|^2$ is used. The displayed quantities are for on-axis vector potentials ($\rho = 0$). Black line: $n = 1$, blue line: $n = 3$, and red-dashed line: $n = 5$. For all lines, $w_0 = 0.6 \lambda_0$, $L = 0.8 \lambda_0$, $\varphi_0 = 0$, and $n_0 = 5\times10^{25}$ m$^{-3}$. All are snapshots taken at $t = 10$ fs.}
\label{fig3}
\end{figure}

For lack of a more rigorous proof of the dominance of the zeroth-order vector potential and, by extension, the zeroth-order field components to be derived below, the following may be taken as representing a reasonable estimate of the uncertainty in the square of the normalized vector potential, when all corrections to the zeroth-order term have been dropped
\begin{equation}\label{uncert}
	\delta^{(n)} \equiv |a^{(n)}|^2 - |a^{(0)}|^2, \quad a^{(n)} = \frac{A^{(n)}}{a_0}.
\end{equation}
This quantity is shown for a specific example with $n=1, 3$ and 5, in Fig. \ref{fig3}. For this particular situation, the uncertainty is less than 5\%, at worst.

\section{The zeroth-order fields}\label{sec:zeroth}

The zeroth-order vector potential, according to Eq. (\ref{am1}) or, equivalently, Eq. (\ref{analytic}) may be written as
\begin{equation}\label{A0}
  A^{(0)} = A_0 \frac{\exp\left[-\frac{\rho^2}{1+\alpha^2}\right]}{\sqrt{1+\alpha^2}} \frac{\sin(\zeta\Delta k/2)}{\zeta\Delta k/2}
              e^{i\varphi^{(0)}}.
\end{equation}
In (\ref{A0}) the total phase of the zeroth-order vector potential is given by
\begin{eqnarray}\label{phi}
  \varphi^{(0)} &=& \varphi_0+k_0\zeta-\tan^{-1}\alpha+\frac{\alpha\rho^2}{1+\alpha^2}-\frac{\alpha k_p^2w_0^2}{4},\\
            a_0 &=& A_0e^{i\varphi_0};\quad \varphi_0 = \text{constant}.
\end{eqnarray}

Note, at this point, that dependence of $A^{(0)}$ in (\ref{A0}) on the parameters of the plasma, through $k_p$, is solely in the phase $\varphi^{(0)}$. Thus, if one regards the quantity $I^{(0)} \equiv |A^{(0)}|^2$, as a measure of the pulse intensity (following \cite{esarey}) the resulting expression for it here will be identical to that in \cite{salaminPRA92-1}, and all issues discussed, and conclusions arrived at in \cite{salaminPRA92-1}, apply here as well, without any change. 

The measured intensity and power of the pulse, however, are defined in terms of the fields $\bm{E}$ and $\bm{B}$, obtained from the space- and time-derivatives. We now turn attention to the electric and magnetic fields. The first of Eqs. (\ref{EB}) yields
\begin{equation}\label{E}
	\bm{E} = -\frac{\partial\bm{A}}{\partial t}-\frac{c^2}{R}\bm{\nabla}(\bm{\nabla}\cdot\bm{A})-\frac{c^2}{R^2}(\bm{\nabla}\cdot\bm{A})\bm{\nabla}\left(\frac{1}{a}\frac{\partial a}{\partial t}\right).
\end{equation}

Employing cylindrical coordinates ($r, \theta, z$) in this equation, the radial and axial electric field components will be obtained, respectively, from
\begin{equation}
\label{Erform} E_r = -\frac{c^2}{R} \frac{\partial}{\partial r} \left(\frac{\partial A}{\partial z}\right) - \frac{c^2}{R^2}
    \left(\frac{\partial A}{\partial z}\right) \frac{\partial}{\partial r} \left(\frac{1}{a} \frac{\partial a}{\partial t}\right),
\end{equation}
and
\begin{equation}
\label{Ezform} E_z = -\frac{\partial A}{\partial t}- \frac{c^2}{R} \frac{\partial^2 A}{\partial z^2} - \frac{c^2}{R^2}
    \left(\frac{\partial A}{\partial z}\right) \frac{\partial}{\partial z} \left(\frac{1}{a} \frac{\partial a}{\partial t}\right).
\end{equation}
On the other hand, there is only one magnetic field component, polarized azimuthally, and can be derived using
\begin{equation}\label{Bthetaform}
  B_\theta = -\frac{\partial A}{\partial r}.
\end{equation}

Expressions will now be found for the lowest-order terms in the fields, from Eqs. (\ref{A0}) and (\ref{Erform})--(\ref{Bthetaform}). After some tedious algebra, one gets

\begin{widetext}
\begin{equation}
\label{Er}  E_r^{(0)} = E\left(\frac{2\rho}{w_0p}\right)
        \left\{\left[\frac{cQ_2}{R}-\frac{ic^2Q_3}{2z_rpR^2}\right]\text{sinc}\left(\frac{\pi\zeta}{L}\right)+\frac{2}{\zeta}\left[\frac{c}{R}-
        \frac{ic^2}{2z_rpR^2}\right]\cos\left(\frac{\pi\zeta}{L}\right)\right\},
\end{equation}
\begin{equation}
\label{Ez} E_z^{(0)} = E
      \left\{\left[Q_1+\frac{cQ_4}{R}+\frac{c^2Q_3Q_5}{R^2}\right]\text{sinc}\left(\frac{\pi\zeta}{L}\right) +\frac{2}{\zeta}\left[1-\frac{cQ_3}{R}+\frac{c^2Q_5}{R^2}\right]\cos\left(\frac{\pi\zeta}{L}\right)\right\},
\end{equation}
\begin{equation}
\label{Btheta} cB_{\theta}^{(0)} = E\left(\frac{4\rho}{w_0p}\right) \text{sinc}\left(\frac{\pi\zeta}{L}\right).
\end{equation}
In Eqs. (\ref{Er})--(\ref{Btheta})
\begin{equation}
    \quad E = \left(\frac{E_0}{2k_0p}\right) \exp\left[i\varphi_0+ik_0\zeta-\frac{\rho^2}{p}-\frac{i\alpha k_p^2w_0^2}{4}\right];\quad E_0 = ck_0a_0;\quad R = \frac{c}{2}\left[Q_1 + \frac{2\pi}{L} \cot\left(\frac{\pi\zeta}{L}\right)\right],
\end{equation}
\begin{equation}
   Q_1 = 2ik_0-\frac{2}{\zeta}+\frac{i(p-\rho^2)}{z_r p^2}+\frac{ik_p^2}{2k_0}; \quad
   Q_2 = 2ik_0-\frac{2}{\zeta}-\frac{i(2p-\rho^2)}{z_r p^2}-\frac{ik_p^2}{2k_0}; \quad
   Q_3 = 2ik_0-\frac{2}{\zeta}-\frac{i(p-\rho^2)}{z_r p^2}-\frac{ik_p^2}{2k_0},
\end{equation}

\begin{equation}
   Q_4 = 2 k_0^2 + \frac{2\pi^2}{L^2} + \frac{2Q_3}{\zeta} -\frac{2k_0(p-\rho^2)}{ z_rp^2}+\frac{2 p^2-4 p \rho^2 + \rho^4}{2z_r^2 p^4}-k_p^2\left(1-\frac{k_p^2}{8k_0^2}\right)+\frac{k_p^2(p-\rho^2)}{2k_0z_rp^2},
\end{equation}
\begin{equation}
  Q_5 = \frac{1}{\zeta^2} + \frac{p - 2\rho^2}{4z_r^2p^3} - \frac{\pi^2}{L^2}\csc^2\left(\frac{\pi\zeta}{L}\right).
\end{equation}
For definiteness in the applications, the constant amplitude $E_0$ will be replaced by $NE'_0 $, where
\begin{equation}\label{N}
	N = \frac{3 i \left(4k_0^2 z_r+2 k_0+k_p^2 z_r\right)^2}{4 \left[2\pi^2(k_p^2 z_r^2+2k_0 z_r) /L^2-3(8 k_0^3 z_r+4 k_0^2 k_p^2 z_r^2 +2 k_0^2 + 4 k_0 k_p^2 z_r + k_p^4z_r^2)\right]},
 \end{equation}
 \end{widetext}
 is a dimensionless normalization constant guaranteed to make $|E|^2 = |E_0|^2$ at the focal point ($x=y=z=0$) at $t=0$. 

\begin{figure}[b]
\includegraphics[width=8cm]{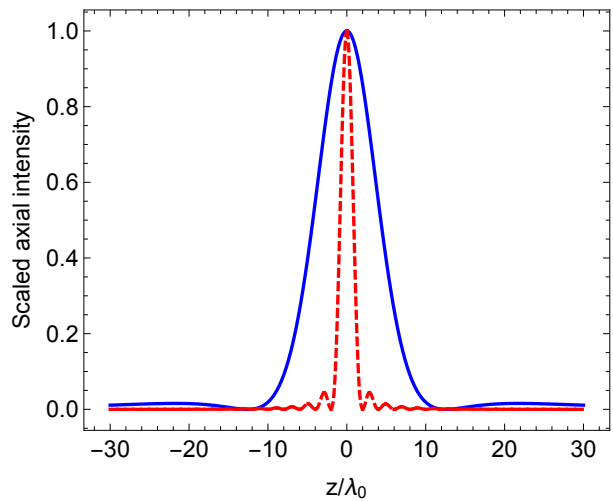}
\caption{(Color online) The scaled axial ($\rho = 0$) intensity profile $|E_r/E'_0|^2$, in a vacuum. Blue line: the paraxial approximation, $w_0 = 2\lambda_0$ and $\varphi_0 = 0$. Red-dashed line: the zeroth-order approximation (this work) at $t = 0$, for $w_0 = L = 2\lambda_0$ and $\varphi_0 = 0$.}
\label{fig4}
\end{figure}

As can be seen quite clearly in Eqs. (\ref{Er})--(\ref{Btheta}) the radial electric field component, $E_r^{(0)}$, and the {\it azimuthal} magnetic field, $B_{\theta}^{(0)}$, vanish identically on the propagation axis ($\rho = 0$), whereas the axial electric field component, $E_z^{(0)}$, is quite strong along the same axis. The former observation gives rise to the well known ring-shaped uniform intensity pattern in a plane transverse to the direction of propagation, while the latter accounts for the well known pencil-like focus along that axis. Both features can be exhibited by, e.g., a Gaussian beam submitted to an axicon. These observations are what make a radially polarized beam, or pulse, well-suited for some important applications, including particle laser acceleration \cite{salamin5} and material processing \cite{meier}.

\begin{figure}[b]
\includegraphics[width=8cm]{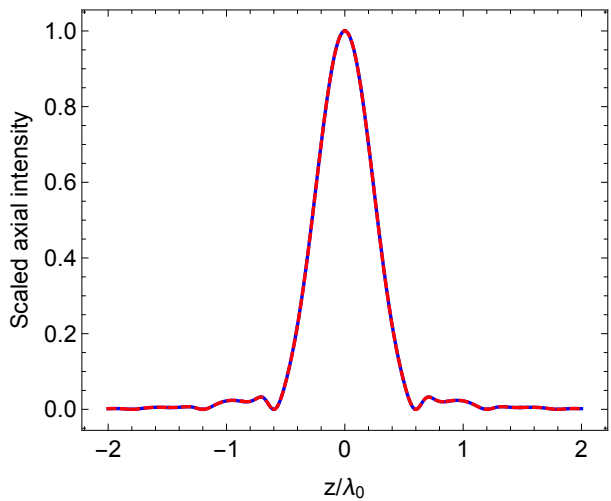}
\caption{(Color online) Initial ($t = 0$) scaled axial ($\rho = 0$) intensity profile $|E_z/E'_0|^2$ of the pulse with parameters $w_0 = L = 0.6 \lambda_0$ and $\varphi_0 = \pi/2$. Blue line: in vacuum ($n_0 = 0$). Red-dashed line: in a plasma of ambient electron number density $n_0 = 5\times10^{25}$ m$^{-3}$. }
\label{fig5}
\end{figure}

\begin{figure}[b]
\includegraphics[width=8cm]{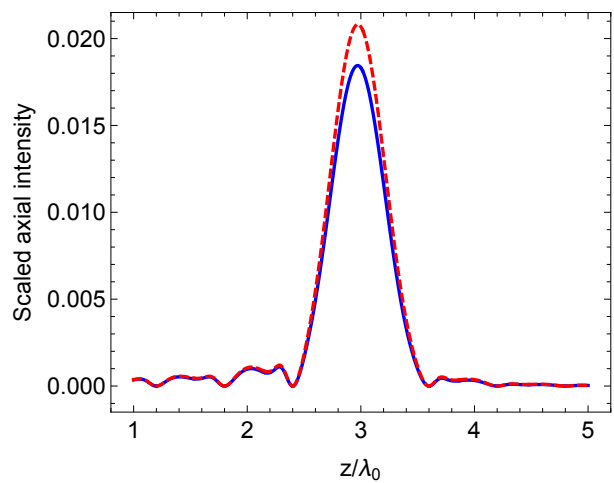}
\caption{(Color online) Scaled axial ($\rho = 0$) intensity profile $|E_z/E'_0|^2$ of the pulse of Fig. \ref{fig5}, after propagation for 10 fs. Blue line: in vacuum ($n_0 = 0$). Red-dashed line: in a plasma of ambient electron number density $n_0 = 5\times10^{25}$ m$^{-3}$. }
\label{fig6}
\end{figure}

The field expressions are quite complicated, even at the zeroth-order level. Unfortunately, our field expressions cannot be directly compared to others arrived at using approaches \cite{april1,april2,varin1,fortin,payeur} different from the one employed here. Numerical comparisons may, in principle be carried out, but will not be done in this paper. Comparing ours analytically even with the corresponding paraxial approximation fields is not straightforward. In the paraxial approximation, the axial field component of a radially polarized beam, with a stationary focus at the origin of coordinates, is given in the literature \cite{esarey4} by
\begin{eqnarray}\label{pa}
	E_z &=& \frac{2N'E_0w_0}{k_0w^2} \exp\left[-\frac{r^2}{w^2} \right]\nonumber\\
	& &\times \left[\left(1-\frac{r^2}{w^2}\right)\cos\psi-\frac{zr^2}{z_rw^2} \sin\psi\right],
\end{eqnarray}
where $w=w_0\sqrt{1+z^2/z_r^2}$, $\psi = \varphi_0+k_0\zeta+zr^2/(z_rw^2)-2\tan^{-1}(z/z_r)$, and $N' = k_0w_0/2$ plays a role similar to $N$ given by Eq. (\ref{N}). Axial intensity profiles based upon Eqs. (\ref{Ez}) and (\ref{pa}) are shown in Fig. \ref{fig4}, employing the same parameters. The obvious conclusion from this figure is that the zeroth-order field (this work) is far better than the dipole approximation in describing $E_z$ for an ultrashort pulse. The zeroth-order approximation profile has a full-width-at-half-maximum of about $2\lambda_0\sim L = w_0$, whereas that of the paraxial approximation profile is roughly $10\lambda_0$.

\section{Pulse propagation}\label{sec:prop}

The zeroth-order field expressions in a plasma (this paper) have been written using the notation and structure introduced in \cite{salaminPRA92-1} for propagation in a vacuum. Clearly, the results of this work have the correct vacuum limits, as $k_p\to0$.  

\begin{figure}[t]
\includegraphics[width=8cm]{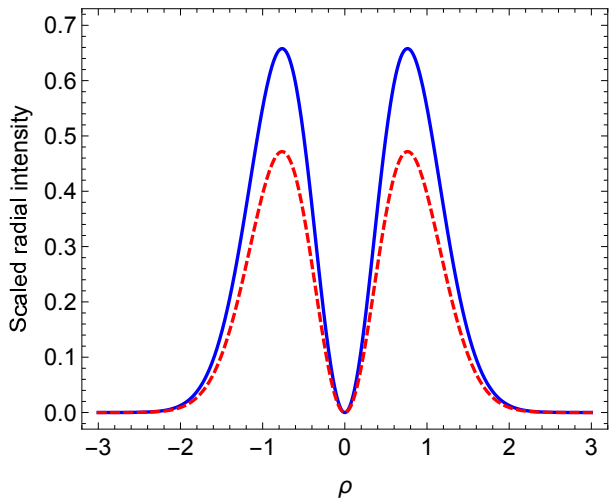}
\caption{(Color online) Initial ($t = 0$) scaled radial ($z = 0$) intensity profile $|E_r/E'_0|^2$ of the pulse with parameters $w_0 = L = 0.6 \lambda_0$ and $\varphi_i = \pi/2$. Blue line: in vacuum ($n_0 = 0$). Red-dashed line: in a plasma of ambient electron number density $n_0 = 5\times10^{25}$ m$^{-3}$. }
\label{fig7}
\end{figure}

\begin{figure}[t]
\includegraphics[width=8cm]{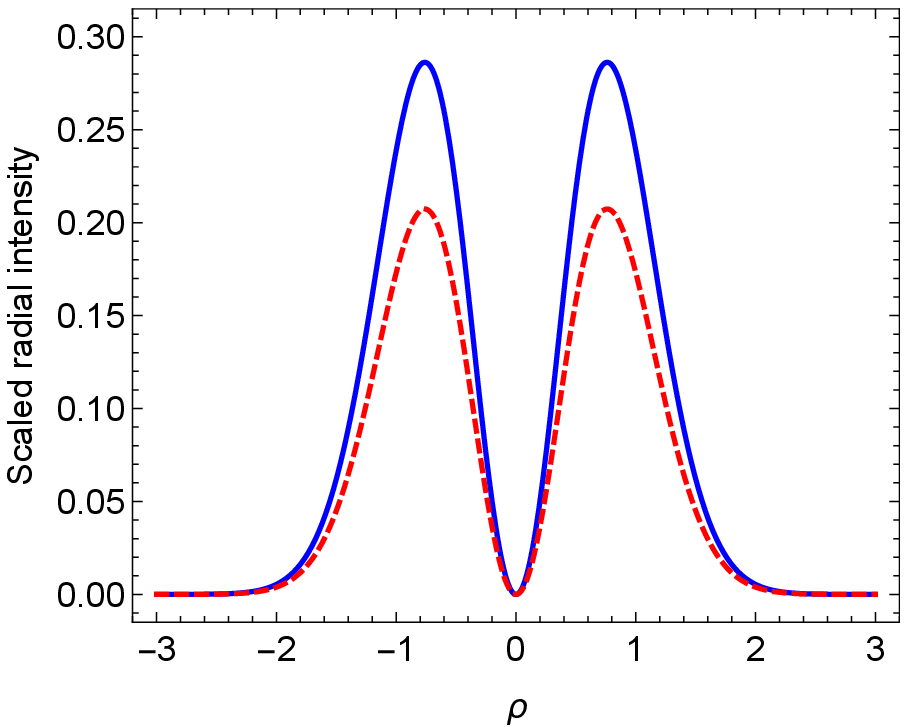}
\caption{(Color online) Scaled radial ($z = 0$) intensity profile $|E_r/E'_0|^2$ of the pulse of Fig. \ref{fig7}, after propagation for 1 fs. Blue line: in vacuum ($n_0 = 0$). Red-dashed line: in a plasma of ambient electron number density $n_0 = 5\times10^{25}$ m$^{-3}$. }
\label{fig8}
\end{figure}

Based on the zeroth-order fields alone, propagation of an ultrashort, tightly focused, radially polarized laser pulse in a plasma, begins to be altered appreciably, from the vacuum case, only for plasma frequencies roughly in excess of $6\times10^{12}~ s^{-1}$, which corresponds to an ambient electron density $n_0 > 10^{22}~ m^{-3}$. This is demonstrated here by concrete examples in Figs. \ref{fig5}--\ref{fig8}, which display axial and radial intensity profiles for a pulse in vacuum and in the presence of a plasma whose ambient electron number density is $n_0 = 5\times10^{25}$ m$^{-3}$. Each graph is just a snapshot of the profile in question either at $t = 0$ or at a later time. Due to the fact that the fields undergo fast and complex space-time oscillations, it must be stated that these snapshots have been carefully chosen to bring out the qualitative features of propagation and distortion of the pulse. 

For example, Fig. \ref{fig5} shows that the presence of the plasma background does not alter the initial axial profile (before onset of any laser-plasma interaction). Allowing the pulse to propagate for 10 fs, however, results in distortion of the profile.  Figure \ref{fig6} shows a higher peak for the plasma-based profile than if propagation is in vacuum, in this particular snapshot. It also displays an overall sharp decrease in both peak intensities, due to diffraction. 

Note also that the centroid of the profile has advanced a distance of $\Delta z \lesssim 3\lambda_0$ in 10 fs, due to the fact that the pulse propagates along the $z-$axis at a group velocity slightly less than the speed of light in vacuum. This point can be appreciated by investigating the total phase of the zeroth-order vector potential. Consider, for example, the effective pulse frequency, $\omega^{(0)}$, and axial wavenumber, $k_z^{(0)}$, found from \cite{esarey}
\begin{eqnarray}
\label{omega0}  \omega^{(0)} & = & -\frac{\partial\varphi^{(0)}}{\partial t} = ck_0(1+\epsilon), \\
\label{kz0}  k_z^{(0)}  &=& \frac{\partial\varphi^{(0)}}{\partial z} = k_0(1-\epsilon),
\end{eqnarray}
where
\begin{equation}\label{epsilon}
  \epsilon = \frac{k_p^2}{4k_0^2}+\frac{1-\rho^2+\alpha^2(1+\rho^2)}{k_0^2w_0^2\left(1+\alpha^2\right)^2}.
\end{equation}
The following dispersion relation may be obtained from Eqs. (\ref{omega0}) and (\ref{kz0})
\begin{equation}\label{dispersion}
  \left[\frac{\omega^{(0)}}{c}\right]^2-\left[k_z^{(0)}\right]^2 = 4\epsilon k_0^2 .
\end{equation}
Employing the dispersion relation (\ref{dispersion}) it can be shown that the product of the phase and group velocities is $v_p^{(0)} v_g^{(0)} = c^2$, where $v_p^{(0)} = \omega^{(0)}/k_z^{(0)}$ and $v_g^{(0)} = d\omega^{(0)}/dk_z^{(0)}$. Note that $v_p^{(0)} = c(1+\epsilon)/(1-\epsilon) > c$, implying that the group velocity $v_g^{(0)} $, at which energy and information get transported in the medium, is less than the speed of light in vacuum.

Figures \ref{fig7} and \ref{fig8} are snapshots of the radial intensity profile in the focal plane ($z = 0$) taken at $t = 0$ and at $t = 1$ fs, respectively. The effect of the plasma background is apparent in both figures. In these snapshots, the instantaneous space- and time-oscillations are such as to make the profile maxima with the plasma lower than in vacuum.\\

\section{Concluding remarks}\label{sec:conc}

Ionization of the atoms of a target, for example the fuel pellet in a laser fusion experiment, is the first thing that happens to it when hit by a high-power laser pulse, quickly turning the target into a plasma. Such laser-matter interactions can be highly nonlinear. Under conditions of low enough ambient electron densities, the plasma response can be considered approximately linear. Such interactions are best studied numerically, employing particle-in-cell (PIC) codes, for example. The work presented in this paper, however, has been concerned with analytically modelling the electromagnetic fields of an ultrashort and tightly focused radially polarized laser pulse, propagating in a plasma whose response may be considered approximately linear. The hope is that the derived equations will be useful for applications and can help to better understand the interactions involved. In addition to the field equations, Eq. (\ref{dispersion}) may be useful in investigations of the plasma dispersive effects associated with interaction with the laser pulse.

Even the equations giving the lowest-order (most dominant) fields turn out to be a bit complex. They have been written out in terms of a convenient set of coordinates. The transverse coordinate, $\rho$, is the cylindrical $r$ scaled by $w_0$, the waist radius at focus. The other coordinates are $\zeta$ (giving the $z$-coordinate of a point within the pulse relative to its centroid) and $\eta$ (which gives the same relative to the initial focus at $t = 0$).  For a point traveling with the pulse at a speed less than the speed of light in vacuum, for which $z$ is less than $ct$, $\zeta$ is negative (but never zero). For the same point, $\eta < ct$ \cite{esarey}. 

To ensure total compatibility with the full set of Maxwell equations \cite{jackson}, a scalar potential has been employed, along with the vector potential. Overall, the analysis has been complicated further by inclusion of the scalar potential from the outset. A small number of rather involved terms in the reported field expressions can, in principle, be traced back easily to that potential. Their effect, however, is only minor.

\section*{acknowledgments}

Support for this work has been provided partly by the Alexander von Humboldt Stiftung, during a research re-invitation (Wiederaufnahme) at Max Planck Institute for Nuclear Physics in Heidelberg, Germany, and partly by the American University of Sharjah, United Arab Emirates, through a Faculty Research Grant (FRG17-T-14). The author also acknowledges, with thanks, fruitful discussions with Christoph Keitel, Meng Wen, and Naveen Kumar at Max Planck Institute for Nuclear Physics in Heidelberg, Germany.

\end{document}